\documentclass[amsmath,amssymb,aps,nofootinbib,preprint,prd,superscriptaddress]{revtex4-1}

\usepackage{aas_macros,esint,graphicx,orcidlink,tikz}
\usetikzlibrary{angles,arrows.meta,decorations.pathreplacing,positioning,quotes}

\let\originalleft\left
\let\originalright\right
\renewcommand{\left}{\mathopen{}\mathclose\bgroup\originalleft}
\renewcommand{\right}{\aftergroup\egroup\originalright}

\newcommand{\ab}[1]{\left|#1\right|}

\newcommand{\br}[1]{\left[#1\right]}
\newcommand{\cu}[1]{\left\{#1\right\}}
\newcommand{\pa}[1]{\left(#1\right)}

\newcommand{\ed}{\mathop{}\!\mathrm{d}}

\begin{document}

\title{Black Hole Vision:\texorpdfstring{\\}{}
An Interactive iOS Application for Visualizing Black Holes}

\author{Roman Berens\,\orcidlink{0000-0003-1509-5463}}
\email{roman.berens@vanderbilt.edu}
\affiliation{Department of Physics \& Astronomy, Vanderbilt University}

\author{Dominic O. Chang\,\orcidlink{0000-0001-9939-5257}}
\email{dominicchang@fas.harvard.edu}
\affiliation{Department of Physics, Harvard University}

\author{Trevor Gravely\,\orcidlink{0009-0004-6058-843X}}
\email{trevor.gravely@vanderbilt.edu}
\affiliation{Department of Physics \& Astronomy, Vanderbilt University}

\author{\\Alexandru Lupsasca\,\orcidlink{0000-0002-1559-6965}}
\email{alexandru.v.lupsasca@vanderbilt.edu}
\affiliation{OpenAI}
\affiliation{Department of Physics \& Astronomy, Vanderbilt University}

\begin{abstract}
The Black Hole Explorer (BHEX) is a proposed mission to launch a sub-millimeter radio telescope into Earth orbit that will take the sharpest images in the history of astronomy and reveal novel horizon-scale features of supermassive black holes.
Black Hole Vision\footnote{Source code available at \url{https://github.com/graveltr/BlackHoleVision}.} is an open-source application, freely available on the iOS App Store, that produces lensed images which highlight the key features expected to appear in the black hole images BHEX will capture.
The app combines video feeds from the front- and rear-facing iPhone cameras and uses the black hole lensing equations to synthesize an onscreen image displaying the user's surroundings as if they were gravitationally lensed by a black hole within the cameras' field of view.
Here, we describe how light rays are lensed by non-rotating (Schwarzschild) and rotating (Kerr) black holes, and we list the equations needed for computing black-hole-lensed images.
We also describe their specific implementation within Black Hole Vision.
\end{abstract}

\maketitle

\section{Introduction}

In a 2019 breakthrough, the Event Horizon Telescope (EHT) collaboration released the first images of M87*, the supermassive black hole at the center of the galaxy M87 \cite{EHT2019a}.
These were followed in 2022 by the first images of Sagittarius A* \cite{EHT2022a}, the supermassive black hole at the center of our own galaxy, the Milky Way. 
The Black Hole Explorer (BHEX) is a proposed mission to take the sharpest images in the history of astronomy by extending the EHT ground array to space with an orbiting telescope \cite{BHEX2024a}.
BHEX will deliver high-resolution horizon-scale images of M87* and Sgr A*, resolving the light rays that are highly bent by the strong gravity of these sources and probing detailed properties of the spacetime geometry near their event horizons \cite{BHEX2024b,BHEX2024c}. 
These images will reveal novel phenomena, including the ``photon ring'': a sharp, bright ring of light predicted (but not yet observed \cite{Gralla2022,Tiede2022}) by general relativity to appear in black hole images due to the existence of nearly bound photons that orbit the event horizon before eventually escaping to our telescopes \cite{GrallaLupsasca2020a,GrallaLupsasca2020b}. 
BHEX will confirm this striking prediction of Einstein's theory and advance our understanding of black holes.

Given the groundbreaking success of the EHT, and with BHEX now on the horizon, there is increasing interest in the rapidly growing field of black hole imaging. 
This is therefore an opportune time to release \textit{Black Hole Vision} (BHV): an open-source, interactive application for iOS that visually demonstrates the effects of black hole lensing upon a user's immediate surroundings. The app captures live video feeds from the front- and rear-facing cameras of the iPhone and uses the black hole lensing equations to generate a physically accurate, lensed image of the cameras' field of view. 
The resulting view displays key features of black hole images, including the photon ring effect targeted by BHEX, and we hope that the app will provide its users with a compelling demonstration of strong gravitational lensing. 
Another aim of this app is to stoke enthusiasm for black hole physics---and for the field of general relativity more generally---by introducing new students to the black hole lensing problem.

In this paper, we explain the physics underlying BHV and provide an introduction to the relevant literature on black hole lensing.
The paper is structured as follows. 
In Sec.~\ref{sec:BlackHoleVision}, we discuss the physical setup implemented within the app and provide an overview of the procedure used to generate its output.
Then in Sec.~\ref{sec:SchwarzschildLensing}, we consider lensing by a non-rotating (Schwarzschild) black hole, before generalizing in Sec.~\ref{sec:KerrLensing} to the rotating (Kerr) case. 

\newpage

\section{Black Hole Vision}
\label{sec:BlackHoleVision}

Before discussing the detailed mathematics of light bending in a black hole spacetime, we begin with a brief discussion of the physical setup implemented in BHV and provide an overview of the procedure that must be followed to generate its output. 

In BHV, we consider a rotating (Kerr) black hole of mass $M$ and angular momentum $J=aM$, where $a$ is usually referred to as the spin parameter.
The black hole is centered at the origin with its spin axis pointing upwards and is observed from a large distance $r_{\rm o}\gg M$.
A non-rotating (Schwarzschild) black hole, which has spin $a=0$, is spherically symmetric and therefore appears the same to all observers in any direction $(\theta_{\rm o},\phi_{\rm o})$.
A nonzero spin breaks spherical symmetry down to axisymmetry about the axis of rotation.
Hence, when $a\neq0$, we can still position the observer at $\phi_{\rm o}=0$ without loss of generality, but now the appearance of the black hole does depend on the observer inclination $\theta_{\rm o}\in [0,\pi/2)$ with respect to the spin axis.
Since iPhone camera images lack depth (that is, they do not relay information about the distance of objects within the field of view), we make the simplifying assumption that all photons captured by the observer were emitted from a ``source sphere'' of constant radius $r_{\rm s}$.
We endow locations on the source sphere with color by projecting the colored images from the front- and rear-facing cameras onto the two hemispheres of the source sphere defined by the observer's forward and backward fields of view, respectively.

\begin{figure}[ht!]
    \centering
    \includegraphics[width=0.5\linewidth]{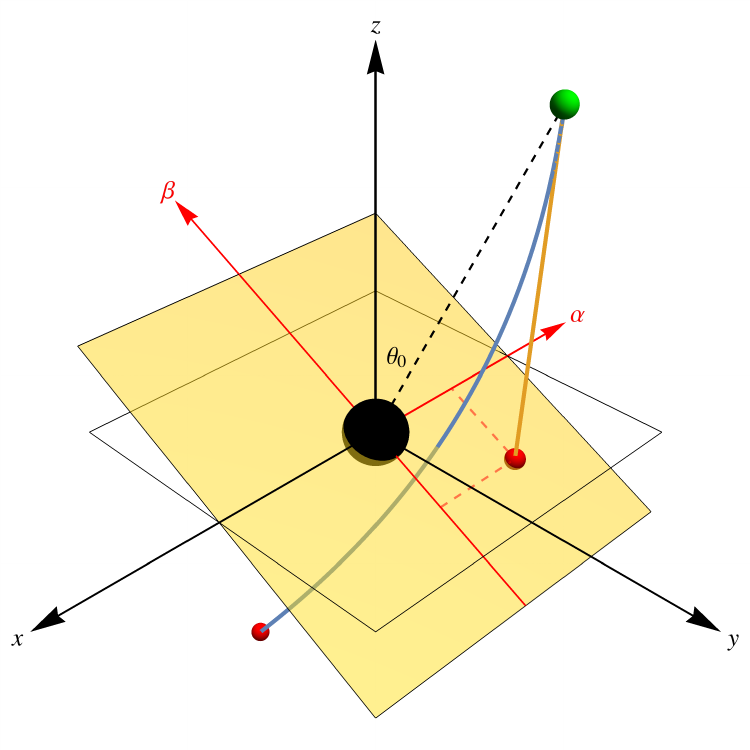}
    \caption{The screen (yellow plane) and Cartesian coordinates $(\alpha,\beta)$ of an observer (green sphere) at large distance from a black hole (black sphere).
    The observer screen is the plane through the black hole that is perpendicular to the observer's line of sight (dashed, black line).
    A photon that reaches the observer traveling along a bent light ray (blue arc) produces an image on the screen (red sphere within the yellow plane) at the apparent position of the source in the sky of the observer.
    This figure originally appeared in Appendix E of \cite{Gralla2017}.}
    \label{fig:BardeenCoordinates}
\end{figure}

As illustrated in Fig.~\ref{fig:BardeenCoordinates}, we consider the observer's line of sight to the black hole (dashed, black line) and define our ``observer screen'' to be the (yellow) plane perpendicular to this line of sight that passes through the center of the black hole.\footnote{To make this precise, here we must work in a fictitious flat spacetime obtained by identifying the spherical coordinates on Minkowski spacetime with the Boyer-Lindquist coordinates on the (asymptotically flat) Kerr spacetime; this identification is to be made at large radius, where the two geometries coincide.}
The projection of the spin axis onto this plane defines a $\beta$ axis, which together with a perpendicular axis $\alpha$ defines Cartesian ``screen coordinates'' $(\alpha,\beta)$ on this image plane.
These screen coordinates parameterize the apparent position on the plane of the sky of a light ray reaching the observer.

Photons in the Kerr spacetime carry three conserved quantities: an energy $E=-p_t$, an azimuthal angular momentum $L=p_\phi$, and a Carter constant $Q=p_\theta^2-a^2p_t^2\cos^2{\theta}+p_\phi^2\cot^2{\theta}$.
It will be useful to introduce the ``specific'' (energy-rescaled) quantities
\begin{align}
    \label{eq:MomentumComponentsToConservedQuantities}
    \lambda\equiv\frac{L}{E}
    =-\frac{p_\phi}{p_t},\qquad
    \eta\equiv\frac{Q}{E^2}
    =\frac{p_\theta^2+p_\phi^2\cot^2{\theta}}{p_t^2}-a^2\cos^2{\theta}.
\end{align}
These completely determine the trajectory of a photon (see also, e.g., \cite{Lupsasca2024} for more details).

A light ray shot backwards from the observer into the geometry in the direction $(\alpha,\beta)$ has conserved quantities of motion
\begin{align}
    \label{eq:ScreenCoordinatesToConservedQuantities}
    \lambda=\alpha\sin{\theta_{\rm o}},\qquad
    \eta=\pa{\alpha^2-a^2}\cos^2{\theta_{\rm o}}+\beta^2.
\end{align}
Thus, we see that a choice of screen coordinates $(\alpha, \beta)$ picks out a momentum of a photon emanating from our observer.\footnote{We consider photons emanating from the observer screen rather than impinging upon it because deducing where a light ray with screen coordinates $(\alpha,\beta)$ was emitted from requires evolving such a ray \textit{backwards} in time from the observer to the source sphere.
This is what we implement in BHV.}

Ultimately, BHV produces a lensed image that is displayed on the iPhone screen.
This image is a matrix of pixels, and each pixel is endowed with a color.
We can label every pixel in the matrix with Cartesian ``pixel coordinates'' $(i, j)$, where for convenience we let the origin of these pixel coordinates correspond to the central pixel in our matrix.

To generate the BHV lensed image, we must determine the color of all its pixels.
For a pixel with coordinates $(i,j)$, we determine its color by implementing the following steps:
\begin{enumerate}
    \item We map the pixel coordinates $(i,j)$ to screen coordinates $(\alpha,\beta)=(\alpha(i,j),\beta(i,j))$.     
    \item We map $(\alpha,\beta)$ to conserved quantities $(\lambda,\eta)$ and a momentum $p^\mu$ via Eqs.~\eqref{eq:MomentumComponentsToConservedQuantities}--\eqref{eq:ScreenCoordinatesToConservedQuantities}.
    \item The momentum $p^\mu$, along with the observer's location, define initial conditions for a trajectory in the black hole spacetime.
    We compute the intersection point of this trajectory with the source sphere.
    This intersection point on the source sphere has a color (determined from the outset by the stretching of the front- and rear-facing images onto the source sphere hemispheres).
    Finally, our lensed image pixel with coordinates $(i,j)$ is assigned this color.
\end{enumerate}
Repeating the procedure for each pixel in the lensed image produces the output of BHV.

To convert pixel coordinates $(i,j)$ to screen coordinates $(\alpha,\beta)$ and thereby implement step (1), we use a linear scaling such that 
\begin{align}
    \label{eq:PixelCoordinatesToScreenCoordinates}
    (\alpha,\beta)=\gamma(i,j),
\end{align}
for some scale factor $\gamma\in\mathbb{R}$.
This is equivalent to uniformly stretching the matrix of pixels that constitute the lensed image onto the screen in Fig.~\ref{fig:BardeenCoordinates}.
The choice of $\gamma$ is arbitrary, and the value we select in BHV is determined purely by aesthetic preference.
This completes the discussion of steps (1) and (2).  

Step (3) consists of mapping conserved quantities $(\lambda,\eta)$ or equivalently a momentum $p^\mu$ to a location on the source sphere.
We defer the detailed calculations involved in step (3) to Secs.~\ref{sec:SchwarzschildLensing} and \ref{sec:KerrLensing}.
For the remainder of this section, we restrict our attention to the special case of a non-rotating Schwarzschild black hole, in which the problem simplifies considerably.

In Schwarzschild, we may use the rotational symmetry of the spacetime to place the observer in the equatorial plane, such that $\theta_{\rm o}=\pi/2$ (recall that we already set $\phi_{\rm o}=0$).
Moreover, thanks to rotational symmetry, each photon trajectory lies within a single plane.
More precisely, consider a ray of constant polar angle $\psi$ on the screen of Fig.~\ref{fig:BardeenCoordinates} (for our equatorial observer, this screen is the $yz$-plane).
All the trajectories emanating from the observer with an apparent position on this ray of angle $\psi$ will lie in a single plane of motion.\footnote{For our equatorial observer, this is simply the $xz$-plane rotated by $\psi$ about the $x$-axis.}
In addition, due to our choice of mapping from pixel coordinates to screen coordinates defined in Eq.~\eqref{eq:PixelCoordinatesToScreenCoordinates}, a ray of constant polar angle $\psi$ on the screen corresponds to a ray of constant polar angle $\psi$ in the $(i,j)$-plane of our lensed image.
Thus, we can decompose our lensed image into rays of constant $\psi$ and then work one ray at a time, with all subsequent lensing calculations taking place within a single plane of motion. 

\begin{figure}[b!]
    \centering
    \begin{tikzpicture}[scale=3.0]
    
    \coordinate (A) at (-1, -1);
    \coordinate (B) at (0, 0);
    \coordinate (C) at (1.5, 0);
    \coordinate (D) at (0, 0.8);
    \draw[black, dashed] (A) -- (B) node[midway, right, xshift=0.1cm] {$r_{\rm s}$};
    \draw[black, dashed] (B) -- (C) node[midway, below] {$r_{\rm o}$};
    \draw[blue, thick] (C) to[out=145, in=85] (A);
    \pic [draw, angle eccentricity=1.2, angle radius=0.7cm] {angle = C--B--A};
    \node at (-0.3, 0.25) {$\phi_S(\lambda)$};
    \draw [black, dashed] (0,0) 
        ++(120:1.4142) 
        arc[start angle=120, end angle=240, radius=1.4142];
    \draw[black, dashed] (0, 1.2) -- (0, -1.2);
    \draw[orange] (C) -- (0, 1.05);
    \draw[black] (0, 0) -- (0, 1.05);
    \node at (-0.1, 0.5) {$\lambda$};
    \draw[fill=black] (B) circle (2pt);
    \draw[fill=red] (A) circle (0.8pt);
    \draw[fill=green] (C) circle (0.8pt);
    \draw[fill=red] (0, 1.05) circle (0.8pt);
    \node at (-1.625, 0.84) {Source Sphere};
    \node at (0.25, -1.0) {Screen};

    \end{tikzpicture}
    \caption{In Schwarzschild, for a given choice of polar angle $\psi$ on the screen, coordinates can be chosen such that all photons with apparent position on the ray of constant $\psi$ have trajectories that lie within the equatorial plane---one such trajectory is illustrated here (blue arc).
    To implement step (3), one needs only one lensing calculation: the computation of $\phi_S(\lambda)$, the azimuthal angle accrued along the photon trajectory from observer (green circle) to source sphere (dashed, circular arc).
    The intersection point (red circle) is used to determine the color of the corresponding pixel. Also shown is the apparent position (red circle) on the screen (dashed, vertical line) of the photon trajectory under consideration, with its corresponding distance of $\lambda = \sqrt{\alpha^2 + \beta^2}$.}
    \label{fig:SchwarzschildLensing}
\end{figure}
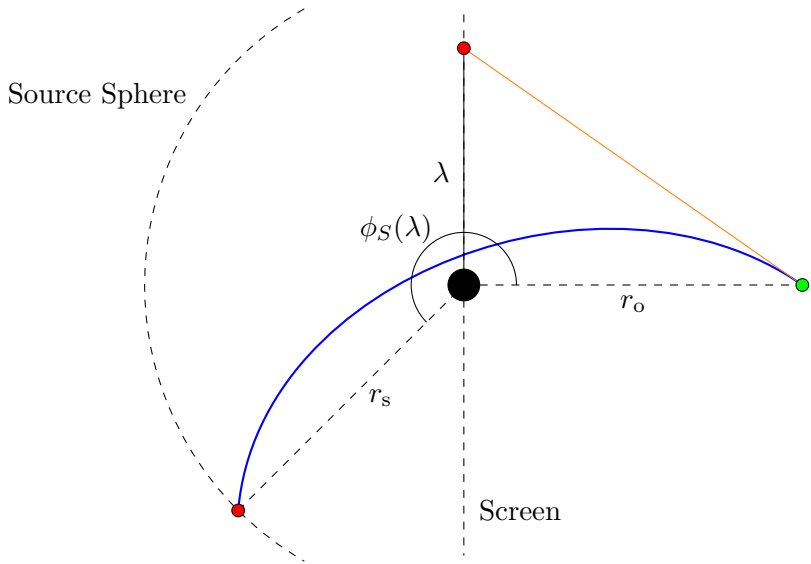

Furthermore, at fixed $\psi$ in the observer screen, we can always rotate our coordinates such that the $\alpha$ axis aligns with the ray of polar angle $\psi$.
Note that under this rotation, all screen positions originally on the ray of constant $\psi$ transform such that $\beta\to0$ and $\alpha\to\sqrt{\alpha^2+\beta^2}$.
Then Eq.~\eqref{eq:ScreenCoordinatesToConservedQuantities} simplifies to
\begin{align}
    \lambda=\sqrt{\alpha^2+\beta^2},\qquad 
    \eta =0,
\end{align}
where we have set $a=0$ (non-rotating black hole) and also used the fact that $\theta_{\rm o} = \pi / 2$. From Eq.~\eqref{eq:MomentumComponentsToConservedQuantities} we see that this sets $p_\theta=0$.
Thus, with this choice of coordinates, the plane of motion becomes the equatorial plane, and $\theta=\pi/2$ along the entire ray.

Under these simplifications, we have reduced the degrees of freedom of photon motion, and the analysis of the photon trajectory takes place entirely within the equatorial plane.
To implement step (3), that is, to compute the position on the source sphere intersected by a trajectory with parameter $\lambda$, we need to compute a single map $\phi_S(\lambda)$.
For an illustration, see Fig.~\ref{fig:SchwarzschildLensing}.
We have thus reduced the entire problem to the computation of a single function $\phi_S(\lambda)$, which encodes all the physics of black hole lensing.
We derive this function in Sec.~\ref{sec:SchwarzschildLensing}.

In the Kerr spacetime, full rotational symmetry is broken and photon trajectories are no longer planar, so one can no longer decompose the image into rays of constant $\psi$.
Instead, the problem is genuinely two-dimensional and one must now work with the full expressions in Eqs.~\eqref{eq:MomentumComponentsToConservedQuantities}--\eqref{eq:ScreenCoordinatesToConservedQuantities}.
Correspondingly, the implementation of step (3) requires the computation of two maps $\phi_K(\lambda,\eta)$ and $\theta_K(\lambda,\eta)$ tracking the azimuthal and polar angles of the intersection point of the trajectory with the source sphere.
We compute these maps in Sec.~\ref{sec:KerrLensing}.

\section{Gravitational Lensing in Schwarzschild}
\label{sec:SchwarzschildLensing}

In this section, we derive the lensing map $\phi_S(\lambda)$ that was defined in Sec.~\ref{sec:BlackHoleVision}, which is needed to compute the output of BHV for the non-rotating (Schwarzschild) black hole.

The spacetime of a non-rotating black hole is described by the Schwarzschild geometry, with line element (throughout, we set $G=c=1$ and work in $\cu{-,+,+,+}$ signature)
\begin{align}
    ds^2=-\pa{1-\frac{2M}{r}}\ed t^2+\pa{1-\frac{2 M}{r}}^{-1}\ed r^2+r^2\pa{\ed\theta^2+\sin^2{\theta}\ed\phi^2}.
\end{align}

As described in Sec.~\ref{sec:BlackHoleVision}, by a change of coordinates we can ensure that any photon remains within the equatorial plane, thereby reducing the degrees of freedom of its geodesic motion and fixing $(\theta,p_\theta)=(\pi/2,0)$ along the entire light ray.
Equatorial trajectories have conserved energy and azimuthal angular momentum
\begin{align}
    E=-p_t,\qquad
    L=p_\phi.
\end{align}
We work with the ``specific'' (energy-rescaled) azimuthal angular momentum
\begin{align}
    \lambda=\frac{L}{E}.
\end{align}
The geodesic equation is then (e.g., \cite{GrallaLupsasca2020b})
\begin{subequations}
\label{eq:SchwarzschildMomentum}
\begin{align}
    \frac{r^2}{E}p^r&=\pm_r\sqrt{\mathcal{R}(r)},\\
    \frac{r^2}{E}p^\phi&=\lambda,\\
    \frac{r^2}{E}p^t&=\frac{r^4}{r(r-2M)},
\end{align}
\end{subequations}
where $\pm_r$ indicates the sign of $p^r$, and we also introduced a radial geodesic potential 
\begin{align}
    \mathcal{R}(r)=r^4-r(r-2M)\lambda^2.
\end{align}
One way to proceed is to define the ``Mino time'' $\tau$ via \cite{Mino2003,GrallaLupsasca2020b}
\begin{align}
    \frac{dx^\mu}{d\tau}=\frac{r^2}{E}p^\mu,
\end{align}
and then use Eqs.~\eqref{eq:SchwarzschildMomentum} to obtain a system of coupled ordinary differential equations for the coordinate motions $x^\mu(\tau)$.
Another method is to convert Eq.~\eqref{eq:SchwarzschildMomentum} into integral form:
\begin{subequations}
\begin{align}
    \tau&=I_r,\\
    \Delta\phi\equiv\phi_{\rm s}-\phi_{\rm o}
    &=I_\phi,\\
    \Delta t\equiv t_{\rm s}-t_{\rm o}
    &=I_t,
\end{align}
\end{subequations}
with 
\begin{subequations}
\begin{align}
    I_r&=\fint_{r_{\rm o}}^{r_{\rm s}}\frac{\ed r}{\pm_r\sqrt{\mathcal{R}(r)}},\\
    \label{eq:Winding}
    I_\phi&=\fint_{r_{\rm o}}^{r_{\rm s}}\frac{\lambda\ed r}{\pm_r\sqrt{\mathcal{R}(r)}}, \\
    I_t&=\fint_{r_{\rm o}}^{r_{\rm s}}\frac{r^4\ed r}{\pm_rr(r-2M)\sqrt{\mathcal{R}(r)}},
\end{align}
\end{subequations}
where the slashed integrals above are understood to be path integrals along the geodesic, while $x_{\rm o}^\mu$ and $x_{\rm s}^\mu$ respectively denote the ``observer'' and ``source'' points.\footnote{Here, the photons travel from the observer at $x_{\rm o}^\mu$ back to the source sphere at $x_{\rm s}^\mu$.
This differs from the notation of \cite{GrallaLupsasca2020b} in which photons travel from source to observer.}

In BHV, the observer radius and source radius are fixed at $r_{\rm o}$ and $r_{\rm s}$, respectively.
In addition, we do not implement time delay.\footnote{Implementing time delay would mean that the output of BHV depends on the history of images captured by the cameras and would thus require images to be cached over time.
Such caching would lead to huge memory consumption, which we opted to avoid (but we include expressions for $\Delta t$ for completeness).}
As a result, we only need to compute the azimuth $\Delta\phi=I_\phi$ swept as the photon travels from $r_{\rm o}$ to $r_{\rm s}$, with $I_r$ given above.

To compute this integral, we further note that in BHV, $r_{\rm o}<r_{\rm s}$ and $p_{\rm o}^r<0$.
That is, photons are launched from some radius $r_{\rm o}$ inwards towards the black hole and exhibit two possible behaviors: either (1) the trajectory intersects the event horizon on the inbound portion of the photon trajectory or (2) the photon falls inward, hits a minimum radius, turns around, and travels all the way back out to $r_{\rm s}$. 

The trajectories in case (1) make up the ``shadow'' of the black hole, which appears as a central dark patch in the output of BHV.
We thus restrict to case (2), in which the radial turning point is the largest root of the radial potential, that is, the largest solution to
\begin{align}
    \mathcal{R}(r)=0.
\end{align}
Letting $b=\ab{\lambda}$, the roots can be obtained via Cardano's method and are given by \cite{Gates2020}
\begin{subequations}
\begin{align}
    r_1&=-\frac{2b}{\sqrt{3}}\cos\br{\frac{1}{3}\arccos\pa{\frac{b_c}{b}}},\\
    r_2&=0,\\
    r_3&=\frac{2b}{\sqrt{3}}\sin\br{\frac{1}{3}\arcsin\pa{\frac{b_c}{b}}},\\
    r_4&=\frac{2b}{\sqrt{3}}\cos\br{\frac{1}{3}\arccos\pa{-\frac{b_c}{b}}},
\end{align}
\end{subequations}
where $b_c=3\sqrt{3}M$.
Case (2) corresponds to $b_c<b$, for which all the above roots are real and ordered as $r_1<r_2<r_3<r_4$.
In BHV, $r_{\rm o}$ is such that $r_4 < r_{\rm o}$, so the turning point encountered by trajectories in case (2) is $r_4$.
Then Eq.~\eqref{eq:Winding} gives
\begin{align}
    \Delta\phi=\lambda\br{\int_{r_4}^{r_{\rm o}}\frac{\ed r}{\sqrt{\mathcal{R}(r)}}+\int_{r_4}^{r_{\rm s}}\frac{\ed r}{\sqrt{\mathcal{R}(r)}}},
\end{align}
where the integrals now denote standard integration in $r$.
These integrals are elliptic, and the relevant antiderivative can be expressed in terms of a known special function as follows.
Define
\begin{align}
    r_{ij}=r_i-r_j,
\end{align}
and let
\begin{align}
    F_i=F\pa{\arcsin\sqrt{\frac{r_{i4}}{r_{i3}}\frac{r_{31}}{r_{41}}}\bigg|\frac{r_{32}r_{41}}{r_{31}r_{42}}},
\end{align}
where $F(\varphi|k)$ denotes the incomplete elliptic integral of the first kind, defined as 
\begin{align}
    F(\varphi|k)=\int_0^\varphi\frac{\ed\theta}{\sqrt{1-k\sin^2{\theta}}}.
\end{align}
One then finds that \cite{GrallaLupsasca2020b}
\begin{align}
    \Delta\phi=\frac{2\lambda}{\sqrt{r_{31}r_{42}}}\br{F_{\rm o}+F_{\rm s}}\label{eq:lensing_solution}.
\end{align}
This is the function $\phi_S(\lambda)=\Delta\phi$ that we need to implement step (3) of the procedure outlined in Sec.~\ref{sec:BlackHoleVision}, which completes the lensing computation for a Schwarzschild black hole.

\section{Gravitational Lensing in Kerr}
\label{sec:KerrLensing}

In this section, we derive the lensing maps $\phi_K(\lambda,\eta)$ and $\theta_K(\lambda,\eta)$ that were defined in Sec.~\ref{sec:BlackHoleVision}, which are needed to compute the output of BHV for the rotating (Kerr) black hole.
The derivation is analogous to that of $\phi_S(\lambda)$ in Sec.~\ref{sec:SchwarzschildLensing}, and closely follows Sec.~II of \cite{GrallaLupsasca2020b}. 

The spacetime of a rotating black hole is described by the Kerr metric, with line element%
\begin{subequations}
\label{eq:Kerr}
\begin{gather}
    ds^2=-\frac{\Delta(r)}{\Sigma}\pa{\ed t-a\sin^2{\theta}\ed\phi}^2+\frac{\Sigma}{\Delta(r)}\ed r^2+\Sigma\ed\theta^2+\frac{\sin^2{\theta}}{\Sigma}\br{\pa{r^2+a^2}\ed\phi-a\ed t}^2,\\
    \Delta(r)=r^2-2Mr+a^2,\qquad
    \Sigma=r^2+a^2\cos^2{\theta},
\end{gather}
\end{subequations}
where $M$ is the black hole mass and $0<a<M$ its spin.
With spherical symmetry broken, trajectories are no longer planar and thus not reducible to equatorial motion.
Nevertheless, thanks to the existence of the Carter tensor and the corresponding conserved Carter integral, geodesic motion is still integrable.
In particular, photons admit the conserved quantities
\begin{gather}
    E=-p_t,\qquad
    L=p_\phi,\qquad
    Q=p_\theta^2-\cos^2{\theta}\pa{a^2p_t^2-\frac{p_\phi^2}{\sin^2{\theta}}},
\end{gather}
which respectively correspond to energy, azimuthal angular momentum about the spin axis, and the Carter integral.
We work with the ``specific'' (energy-rescaled) quantities
\begin{gather}
    \lambda=\frac{L}{E},\qquad
    \eta=\frac{Q}{E^2}.
\end{gather}
The geodesic equation is then (e.g., \cite{GrallaLupsasca2020b})
\begin{subequations}
\begin{align}
    \frac{\Sigma}{E}p^r&=\pm_r\sqrt{\mathcal{R}(r)},\\
    \frac{\Sigma}{E}p^\theta&=\pm_\theta\sqrt{\Theta(\theta)},\\
    \frac{\Sigma}{E}p^\phi&=\frac{a}{\Delta(r)}\pa{r^2+a^2-a\lambda}+\frac{\lambda}{\sin^2\theta}-a,\\
    \frac{\Sigma}{E}p^t&=\frac{r^2+a^2}{\Delta(r)}\pa{r^2+a^2-a\lambda}+a\pa{\lambda-a\sin^2{\theta}}.
\end{align}
\end{subequations}
Here, we have introduced radial and angular geodesic potentials
\begin{subequations}
\begin{align}
    \mathcal{R}(r)&=\pa{r^2+a^2-a\lambda}^2-\Delta(r)\br{\eta+\pa{\lambda-a}^2},\\
    \Theta(\theta)&=\eta+a^2\cos^2{\theta}-\lambda^2\cot^2{\theta}.
\end{align}
\end{subequations}
Conversion to integral form yields (e.g., \cite{GrallaLupsasca2020b})
\begin{subequations}
\begin{align}
    I_r&=G_\theta
    \equiv\tau,\\
    \label{eq:Azimuth}
    \Delta\phi\equiv\phi_{\rm s}-\phi_{\rm o}
    &=I_\phi+\lambda G_\phi,\\
    \Delta t\equiv t_{\rm s}-t_{\rm o}
    &=I_t+a^2G_t, 
\end{align}
\end{subequations}
where $\tau$ is the so-called ``Mino time'' that decouples the radial and polar motions, while
\begin{subequations}
\begin{align}
    I_r&=\fint_{r_{\rm o}}^{r_{\rm s}}\frac{\ed r}{\pm_r\sqrt{\mathcal{R}\pa{r}}},
    &&G_\theta=\fint_{\theta_{\rm o}}^{\theta_{\rm s}}\frac{\ed\theta}{\pm_\theta\sqrt{\Theta\pa{\theta}}},\\
    I_\phi&=\fint_{r_{\rm o}}^{r_{\rm s}}\frac{a\pa{2Mr-a\lambda}}{\pm_r\Delta(r)\sqrt{\mathcal{R}\pa{r}}}\ed r,
    &&G_\phi=\fint_{\theta_{\rm o}}^{\theta_{\rm s}}\frac{\csc^2{\theta}}{\pm_\theta\sqrt{\Theta\pa{\theta}}}\ed\theta,\\
    I_t&=\fint_{r_{\rm o}}^{r_{\rm s}}\frac{r^2\Delta(r)+2Mr\pa{r^2+a^2-a\lambda}}{\pm_r\Delta(r)\sqrt{\mathcal{R}\pa{r}}}\ed r,
    &&G_t=\fint_{\theta_{\rm o}}^{\theta_{\rm s}}\frac{\cos^2{\theta}}{\pm_\theta\sqrt{\Theta\pa{\theta}}}\ed\theta.
\end{align}
\end{subequations}
Here, as in Sec.~\ref{sec:SchwarzschildLensing}, the slashed integrals are understood to be path integrals along the geodesic, with $x_{\rm o}^\mu$ and $x_{\rm s}^\mu$ denoting the ``observer'' and ``source'' points, respectively.

In addition to computing the azimuth swept $\Delta\phi$ in order to obtain the map $\phi_K(\lambda,\eta)$, we must now also compute a map $\theta_K(\lambda,\eta)$ by solving for the final polar angle $\theta_{\rm s}$ that is attained by the photon as it travels from $r_{\rm o}$ to $r_{\rm s}$.
To do so, we compute the photon trajectory parameterized by Mino time, that is, we obtain $\theta(\tau)$ and $\phi(\tau)$ along a light ray leaving $x_{\rm o}^\mu$ with conserved parameters $(\lambda,\eta)$.
More specifically, we compute the total Mino time $\tau_{\rm max}=I_r$ elapsed as the photon travels from $r_{\rm o}$ to $r_{\rm s}$, and then obtain $\phi_K(\lambda,\eta)$ and $\theta_K(\lambda,\eta)$ by evaluating $\Delta\phi(\tau_{\rm max})$ and $\theta_{\rm s}(\tau_{\rm max})$.
As always, we use axisymmetry to set $\phi_{\rm o}=0$.

The integrals above are all elliptic.
We will now provide expressions for them.
We let $\mathcal{I}_i$ and $\mathcal{G}_i$ denote antiderivatives with the choice of a plus sign in the integrand.
For example,
\begin{align}
    \frac{d\mathcal{I}_r}{dr}=\frac{1}{\sqrt{\mathcal{R}(r)}},\qquad
    \frac{d\mathcal{G}_\theta}{d\theta}=\frac{1}{\sqrt{\Theta\pa{\theta}}}.
\end{align}

\subsection{Angular integrals}

We begin by considering the angular integrals; a complete treatment is given in Sec.~III of \cite{GrallaLupsasca2020b}.
First, we define 
\begin{align}
    u_\pm=\Delta_\theta\pm\sqrt{\Delta_\theta^2+\frac{\eta}{a^2}},\qquad
    \Delta_\theta=\frac{1}{2}\pa{1-\frac{\eta+\lambda^2}{a^2}}.
\end{align}
Then direct integration yields \cite{GrallaLupsasca2020b}
\begin{align}
    \mathcal{G}_\theta=-\frac{1}{\sqrt{-u_-a^2}}F\pa{\arcsin\pa{\frac{\cos{\theta}}{\sqrt{u_+}}}\bigg|\frac{u_+}{u_-}}.
\end{align}
Letting $\nu_\theta=\mathrm{sign}\pa{p_{\rm o}^\theta }$ denote the direction of the initial momentum in the $\theta$ direction, we can invert this to obtain the polar trajectory as a function of Mino time $\tau=G_\theta$:
\begin{align}
    \frac{\cos\pa{\theta(\tau)}}{\sqrt{u_+}}=-\nu_\theta\,\mathrm{sn}\pa{\sqrt{-u_-a^2}\pa{\tau+\nu_\theta\mathcal{G}_\theta^{\rm o}}\bigg|\frac{u_+}{u_-}},
\end{align}
where $\mathcal{G}_\theta^{\rm o}$ denotes $\mathcal{G}_\theta$ evaluated at $\theta_{\rm o}$ and $\mathrm{sn}$ is the Jacobi elliptic sine function.
This equation gives $\theta(\tau)$, and therefore $\theta_{\rm s}$ when evaluated at $\tau=I_r$.

Another direct integration yields \cite{GrallaLupsasca2020b}
\begin{align}
    \mathcal{G}_\phi=-\frac{1}{\sqrt{-u_-a^2}}\Pi\pa{u_+;\arcsin\pa{\frac{\cos{\theta}}{\sqrt{u_+}}}\bigg|\frac{u_+}{u_-}},
\end{align}
where $\Pi$ denotes the incomplete elliptic integral of the third kind,\begin{align}
    \Pi(n;\phi|k)=\int_0^\phi\frac{\ed\theta}{\pa{1-n\sin^2{\theta}}\sqrt{1-k\sin^2{\theta}}}.
\end{align}
We further define
\begin{align}
    \Psi_\tau=\mathrm{am}\pa{\sqrt{-u_-a^2}\pa{\tau+\nu_\theta\mathcal{G}_\theta^{\rm o}}\bigg|\frac{u_+}{u_-}},
\end{align}
where $\mathrm{am}$ denotes the Jacobi amplitude: the inverse of the elliptic integral of the first kind,
\begin{align}
    F\big(\mathrm{am}(\varphi|k)\big|k\big)=\varphi.
\end{align}
With these definitions in hand, one can show that \cite{GrallaLupsasca2020b}
\begin{align}
    \label{eq:GphiMino}
    G_\phi=\frac{1}{\sqrt{-u_-a^2}}\Pi\pa{u_+;\Psi_\tau\bigg|\frac{u_+}{u_-}}-\nu_\theta\mathcal{G}_\phi^{\rm o}.
\end{align}
This completes our discussion of the angular integrals.
We have obtained $\theta$ and $G_\phi$, the second term in Eq.~\eqref{eq:Azimuth}, in terms of the elapsed Mino time $\tau$.
The remaining quantities to compute are the radial integral $I_r$, which gives the total Mino time elapsed along the entire trajectory from $r_{\rm o}$ to $r_{\rm s}$, and also the first term in \eqref{eq:Azimuth}, namely, the radial integral $I_\phi$.

\subsection{Radial integrals}

We now consider the radial integrals; for complete details, consult App.~B2 of \cite{GrallaLupsasca2020b}.

To compute the needed radial integrals, we will require the roots of the radial potential.
First, define 
\begin{gather}
    \mathcal{A}=a^2-\eta-\lambda^2,\qquad
    \mathcal{B}=2M\br{\eta+\pa{\lambda-a}^2},\qquad
    \mathcal{C}=-a^2\eta,\\
    \mathcal{P}=-\frac{\mathcal{A}^2}{12}-\mathcal{C},\qquad
    \mathcal{Q}=-\frac{\mathcal{A}}{3}\br{\pa{\frac{\mathcal{A}}{6}}^2-\mathcal{C}}-\frac{\mathcal{B}^2}{8},
\end{gather}
and also
\begin{align}
    \omega_\pm=\sqrt[3]{-\frac{\mathcal{Q}}{2}\pm\sqrt{\pa{\frac{\mathcal{P}}{3}}^3+\pa{\frac{\mathcal{Q}}{2}}^2}},\qquad
    \xi_0=\omega_++\omega_--\frac{\mathcal{A}}{3}.
\end{align}
In terms of $z=\sqrt{\xi_0/2}$, the four roots of the quartic $\mathcal{R}(r)$ are then given by 
\begin{subequations}
\begin{align}
    r_1&=-z-\sqrt{-\frac{\mathcal{A}}{2}-z^2+\frac{\mathcal{B}}{4z}},
    &&r_2=-z+\sqrt{-\frac{\mathcal{A}}{2}-z^2+\frac{\mathcal{B}}{4z}},\\
    r_3&=z-\sqrt{-\frac{\mathcal{A}}{2}-z^2-\frac{\mathcal{B}}{4z}},
    &&r_4=z+\sqrt{-\frac{\mathcal{A}}{2}-z^2-\frac{\mathcal{B}}{4z}}.
\end{align}
\end{subequations}
We will also require the radii of the outer ($r_+$) and inner ($r_-$) horizons of the black hole,
\begin{align}
    r_\pm=M\pm\sqrt{M^2-a^2}.
\end{align}
Finally, we define $r_{ij}\equiv r_i-r_j$ and $r_{\pm i}=r_\pm-r_i$.
We can now tackle the radial integrals.

To begin with, we compute $I_r$: it is given by a sum of its antiderivatives at the endpoints,
\begin{align}
    I_r=\mathcal{I}_r^{\rm o}+\mathcal{I}_r^{\rm s}.
\end{align}
A direct integration shows that \cite{GrallaLupsasca2020b}
\begin{align}
    \label{eq:F2}
    \mathcal{I}_r=F^{(2)}(r)
    \equiv\frac{2}{\sqrt{r_{31}r_{42}}}F\pa{\arcsin{x_2(r)}|k},\qquad
    x_2(r)=\sqrt{\frac{r-r_4}{r-r_3} \frac{r_{31}}{r_{41}}}.
\end{align}
where we introduced the elliptic modulus
\begin{align}
    k=\frac{r_{32}r_{41}}{r_{31}r_{42}}.
\end{align}

Next, we split $I_\phi$ into two integrals:
\begin{subequations}
\begin{align}
    I_\phi&=\frac{2Ma}{r_+-r_-}\br{\pa{r_+-\frac{a\lambda}{2M}}I_+-\pa{r_--\frac{a\lambda}{2M}}I_-},\\
    I_\pm&=\fint_{r_{\rm o}}^{r_{\rm s}}\frac{dr}{\pm_r\pa{r-r_\pm}\sqrt{\mathcal{R}(r)}}.
\end{align}
\end{subequations}
Again, each of the $I_\pm$ is given by a sum of antiderivatives evaluated at the endpoints,
\begin{align}
    I_\pm=\mathcal{I}_\pm^{\rm o}+\mathcal{I}_\pm^{\rm s}.
\end{align}
Another direct integration shows that \cite{GrallaLupsasca2020b}
\begin{align}
    \mathcal{I}_\pm=-\Pi_{\pm}^{(2)}(r)-\frac{F^{(2)}(r)}{r_{\pm 3}},
\end{align}
with $F^{(2)}(r)$ as defined in Eq.~\eqref{eq:F2} and
\begin{align}
    \Pi_\pm^{(2)}=\frac{2}{\sqrt{r_{31}r_{42}}}\frac{r_{43}}{r_{\pm3}r_{\pm4}}\Pi\pa{\frac{r_{\pm3}r_{41}}{r_{\pm4}r_{31}};\arcsin{x_2(r)}\bigg|k}.
\end{align}
This completes the computation of $I_\phi$.
Plugging it into Eq.~\eqref{eq:Azimuth}, along with $G_\phi$ given in Eq.~\eqref{eq:GphiMino}, yields the azimuthal trajectory $\phi(\tau)$ parameterized by Mino time.
Evaluating at $\tau = I_r$ yields the total azimuth swept $\phi_{\rm s}=\phi(I_r)$.

This completes the discussion of lensing in the Kerr spacetime (ignoring time delay).
In summary, we have shown how, given an observer at radius $r_{\rm o}$ and inclination $\theta_{\rm o}$ who shoots a ray backwards into the geometry with conserved quantities $(\lambda,\eta)$, one can explicitly compute the source inclination $\theta_{\rm s}$ and azimuth $\phi_{\rm s}$ of the ray at its source radius $r_{\rm s}$.
This defines the maps $\phi_K(\lambda,\eta)=\phi_{\rm s}$ and $\theta_K(\lambda,\eta)=\theta_{\rm s}$ required in step (3) of the procedure in Sec.~\ref{sec:BlackHoleVision}.

\begin{acknowledgments}

\noindent RB, TG, and AL were supported by the NSF grant AST-2307888 and CAREER award PHY-2340457, and by the Simons Foundation award SFI-MPS-BH-00012593-09.
DC was supported by the Black Hole Initiative, which is funded by grants \#62286 from the John Templeton Foundation and GBMF-8273 from the Gordon and Betty Moore Foundation.

\end{acknowledgments}

\appendix 

\section{Black Hole Mirror}

Black Hole Mirror (BHM) is a precursor to BHV that is available as a \texttt{JavaScript} web application.\footnote{\url{https://dominic-chang.com/bhi-filter/}}
It is computationally lighter than BHV but makes some compromises.
First, to minimize computational cost, BHM only implements the (planar) lensing equations of the Schwarzschild black hole, and not the more complicated ones of the rotating Kerr black hole.
Second, it only uses the feed from the front-facing camera to paint an image on the source sphere, replacing the feed from the rear camera with a star field.
Third, the image from the camera feed is mapped to the source sphere via a gnomonic projection,
\begin{subequations}
\begin{align}
    \theta&=\pi-\arctan\pa{\frac{\sqrt{i^2+j^2}}{r_{\rm s}}}, \\
    \phi&=\mathrm{atan2}(j,i),
\end{align}
\end{subequations}
where $(i, j)$ are the Cartesian pixel coordinates defined in Sec.~\ref{sec:BlackHoleVision}, and where the observer is assumed to be located at the north pole of the coordinate system. 
Since BHM assumes a spherically symmetric spacetime, the analysis of photon trajectories can be done exclusively within the equatorial plane: as explained in Sec.~\ref{sec:BlackHoleVision}, we need only compute the map $\phi_S(\lambda)$. 

Additionally, BHM exaggerates the strength of the lensing effect in Eq.~\eqref{eq:lensing_solution} by including a scaling factor $c>1$.
This scaling maintains the expected image behavior for $\lambda\gg1$.
More specifically, we use a transformed winding angle $\phi_S'(\lambda)$ defined as
\begin{align}
    \phi_S'(\lambda)=c\pa{\phi_S(\lambda)-\pi}+\pi.
\end{align}
This definition ensures that 
\begin{align}
    \lim_{\lambda\rightarrow\infty}\phi'_S(\lambda)=\lim_{\lambda\rightarrow\infty}\phi_S(\lambda)=\pi, 
\end{align}
which is the expected behavior of a photon with large impact parameter.
The numerical value $c=10$ used in BHM is chosen to exaggerate the thickness of the photon-ring subimages for aesthetic appeal.

\bibliographystyle{utphys}
\bibliography{BCGL}

\end{document}